\documentclass[10pt]{article}
\usepackage[margin=1.5in]{geometry}
\usepackage{amsmath}
\usepackage{enumerate}
\usepackage{amssymb}
\usepackage{graphicx}
\usepackage{float}
\usepackage{epstopdf}
\usepackage{color}
\usepackage{ctable}
\usepackage[section,subsection,subsubsection]{extraplaceins}
\usepackage{multirow}
\usepackage{subfigure}
\DeclareMathOperator*{\argmin}{arg\,min}

\begin{document}

\title{\centering \LARGE{Reducing ADC Sampling Rate with Compressive Sensing}}
\author{Morteza Hashemi \\
Electrical and Computer Engineering, Boston University\\
\texttt{mhashemi@bu.edu}\\
 \date{}}
\maketitle

\maketitle
\begin {abstract}

Many communication systems involve high bandwidth, while sparse, radio frequency (RF) signals. Working with high frequency signals requires appropriate system-level components such as high-speed analog-to-digital converters (ADC).  In particular, an analog signal should be sampled at rates that meet the Nyquist requirements to avoid aliasing. However, implementing high-speed ADC devices can be a limiting factor as well as expensive. To mitigate the caveats with high-speed ADC,  the solution space can be explored in several dimensions  such as utilizing the compressive sensing (CS) framework in order to reduce the sampling rate to the order of \emph{information rate} of the signal rather than a rate dictated by the Nyquist. In this note, we review the compressive sensing structure and its extensions for continuous-time signals, which is ultimately used to reduce the sampling rate of high-speed ADC devices. Moreover, we consider the application of the compressive sensing framework in wireless sensor networks to save power by reducing the transmission rate of sensor nodes. We propose an alternative solution for the CS minimization problem that can be solved using gradient descent methods. The modified minimization problem is potentially faster and simpler to implement at the hardware level.    

\end {abstract}

\section{Introduction}\label{sec:Intro}
Considerable advances in Radio Frequency (RF) communications and related 
hardware development make it an appropriate tool for a variety of applications. In 
addition to well-known wireless products, RF technology has other applications 
creating great space for growth. In most of these systems and in order to use off-the-shelf digital signal processing (DSP) blocks, the analog signal first needs to be digitized. However, in order to be able to recover the original signal without aliasing, the sampling rate should meet the Nyquist sampling requirements. In the case of high frequency signals, design and implementation of high-speed analog-to-digital converters (ADC) is a challenge as the Nyquist rate can easily exceed the current capabilities of ADC devices.  For example, some radar applications would be best served by an ADC sampling rate of over 5G samples per second and a resolution of over 20 bits per sample, a 
combination that greatly exceeds current capabilities \cite{kirolos2006analog}. Besides, high-speed ADC devices become very expensive. Within this context, it is necessary to develop algorithms that can relax the requirements for high-speed ADC devices. 

In recent years, compressive sensing (CS) framework has been introduced that enables sparse signals sensing at a sub-Nyquist rate. According to the CS theory, a sparse signal can be represented 
by a number of random projections and recovered from these projections with a high 
probability. The CS framework introduces a novel framework for \emph{analog-to-information} conversion when the target signal is sparse with high frequency components. Therefore, ADC devices can be designed to work based on the Nyquist requirements for the compressed version of signal.

The rest of this report is organized as follows. In Section \ref {sec:sec2}, we provide a brief background on discrete-time compressive sensing, followed by its extension to continuous-time and compressive sensing with parallel branches. Section \ref {rate} focuses on reducing transmission rates in wireless sensor networks using  compressive sensing. Our tweak in the compressive sensing minimization problem is presented in Section \ref{idea}. We conclude with overall thoughts in Section \ref {Conclusion}.

\section{Compressive Sensing}
\label{sec:sec2}
In this section, we review the compressive sensing framework for discrete and continuous time signals, and further explain the application of compressive sensing in ADC sampling rate reduction. 
\subsection{Discrete-time signal}
According to the compressive sensing theory, a discrete-time signal can be represented by a number of 
projections  on a random basis. In particular, given a $K$-Sparse  signal vector $f_{N\times 1}$ and  the  basis 
$ \Psi_{N\times M} $, we have:   
\begin{equation}
f=\Psi \alpha; 
\label{equ1}
\end{equation}
in which $\alpha$ is the coefficient vector for the signal of interest $f$  under the basis $\Psi$.  Here, by 
$K$-Sparse, we mean that $f$ has $K$ nonzero elements only. One key point is that it can be 
proved that sparseness is equivalent to compressibility and vice versa.  
At the receiver side, it is critical to be able to reconstruct the original signal from the compressed version. In this case, assume that an $L\times N $ measurement matrix $\Phi$ and measurement outputs $y_{L\times 1}$  are given such that:
\begin{equation}
y=\Phi f.
\label{equ2}
\end{equation} 
 In this equation, vector $f$  is of interest and needs to be recovered from the matrix $\Phi$  and measurement results  $y$ with the constraint $ L<N$.  
Because Eq. \eqref{equ1} represents an undetermined system of equations, there are infinitely possible solutions for signal. However, if the signal $f$  is known to be sparse, then the most sparse solution would be  acceptable. From \eqref{equ1} and \eqref{equ2}, we have:
\begin{equation}
y=\Phi f=\Phi \Psi\alpha; 
\label{equ3}
\end{equation}
where vector $\alpha$ should be determined. To apply the constraint of sparsity on the vector $\alpha$, we have: 
\begin{equation}
\begin{split}
\alpha^*=\argmin | \alpha |_1, \\
\text{subject to} \ \ y=\Phi \Psi\alpha.
\end{split} 
\end{equation}
In order to solve this minimization problem, there exist several algorithms such as 
Orthogonal Matching Pursuit (OMP)~\cite{tropp2005signal}. Therefore, signal $f$ would be reconstructed as $f^*=\Psi\alpha^*$. 

Before proceeding, we would like to highlight a fact regarding the relationship 
between $\Phi$ and $\Psi$. As discussed earlier, CS relies on the sparseness of the signals. CS 
also depends on the incoherence matrices $\Phi$ and $\Psi$. Fortunately, random sensing 
matrices $\Phi$  are incoherence with any basis and this relaxes knowing basis $\Psi$  a priori. 
Moreover, there are several choices for the distribution of random projections such as 
Gaussian and Bernoulli.

\subsection{Continuous-time signal}
\label{Analog}
\begin{figure}[h]
\centering
\includegraphics[width=9cm]{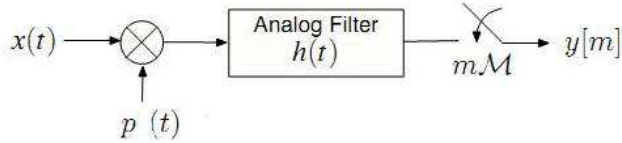}
\caption{CS diagram for analog signals \cite{laska2007theory}}
\label{analogfilter}
\end{figure}

As described in the previous section, compressive sensing framework was first proposed for discrete-time signals. However,  with slight modifications it can be applied to continuous-time signals as well, where  we are interested in using compresisve sensing  to avoid high-speed ADC devices. The authors in \cite{laska2007theory} have proposed a structure shown in Fig. \ref{analogfilter}.

In this structure, the analog signal $x(t)$ is first demodulated by a  pseudo-random sequence  $p(t)$ in order to spread the frequency contents of the signal so that it is not destroyed by the second
stage of the system. In the second step, demodulated signal is filtered by a low-pass filter   $h(t)$,  and finally the signal is sampled by a low-rate ADC. 
Assuming that the signal $x(t)$ is \emph{sparse}, we can decompose it into $N$
basis signals $\psi_i (t)$ as follows:
\begin{equation}
x(t)=\sum_{i=1}^N \alpha_i\psi_i(t). 
\label{equ5} 
\end{equation}
With this representation of signal   $x(t)$     and the block diagram of Fig.\ref{analogfilter} output $y(t)$ becomes:
\newcommand{\ud}{\,\mathrm{d}}
\begin{equation}
y(t)=\int_{-\infty} ^ \infty x(\tau)p(\tau)h(t-\tau) \ud \tau; \label{equ6}
\end{equation} 
and after sampling we have: 
\begin{equation}
y[m]=y(t)|_{t=mM}. \label{equ7}
\end{equation}
By rewriting  $y[m]$    according to the basis expansion of signal  $x(t)$    we have: 
\begin{equation}
y[m]=\sum_{i=1}^N \alpha_i \int_{-\infty} ^ \infty \psi_i(\tau)p(\tau)h(mM-\tau) \ud \tau; \label{equ8}
\end{equation}
and defining: 
\begin{equation}
\nu_{m,i}= \int_{-\infty} ^ \infty \psi_i(\tau)p(\tau)h(mM-\tau) \ud \tau \label{equ9}
\end{equation}
results that: 
\begin{equation}
y[m]=\sum_{i=1}^N \alpha_i \nu_{m,i}. 
\end{equation} 
Or in a matrix format:
$
Y=V\alpha.
$

Note that the matrix $V$  here serves the same purpose of the combination of $\Phi \Psi$ in Eq.\ref{equ3}. 
Therefore, analog signal reconstruction leads to finding an optimal solution for $\alpha$  such that:
\begin{equation}
\begin{split}
\alpha^*=\argmin|\alpha|_1  \\
\text{subject to} \ \ Y=V\alpha. 
\label{equ12}
\end{split}
\end{equation}
The reconstructed  signal is then:
\begin{equation}
x^*(t)=\sum_{i=1}^N \alpha^*_i \psi_i(t) \label{equ13}
\end{equation}                   
where $\alpha^*$ is the solution for 
the optimization problem. Therefore, the compressive sensing structure allows the sparse signal $x(t)$ (i.e., output of the low-pass filter) to be sampled at a sub-Nyquist rate.    

\subsection{Parallel structure}
\label{Parallel Structure}
In the previous section, we studied a serial structure to apply compressive sensing on 
continuous-time signals. Although it is a promising approach to reduce the sampling rate of high-rate ADC devices, to achieve a high-quality signal reconstruction the 
order of filter is typically higher than $10$, which may not be desirable to design and implement complex high-order filers.  To this end, a \emph{parallel segmented compressive sensing (PSCS)} has been proposed in \cite{yu2008mixed}. In this scheme, the continuous signal is first 
segmented into overlapping partitions. Next, compressive sensing is applied to each segment independently, and the outputs of each segment are combined together in order to reconstruct the signal 
efficiently. PSCS can be used in applications such as cognitive radio wherein 
the spectrum sensing is crucial for successful operations.
  
\begin{figure}[tb]
\centering
\includegraphics[width=9cm, trim=0cm 2cm 1cm 0cm, clip]{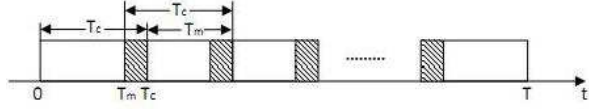}
\caption{Time-windowed analog signal \cite{yu2008mixed}}
\label{window}
\end{figure}

\begin{figure}[tb]
\centering
\includegraphics[width=9cm]{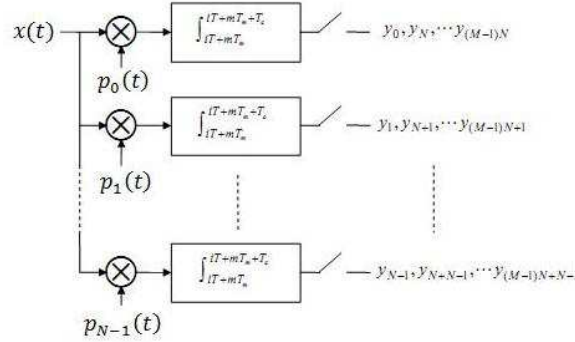}
\caption{Parallel structure of CS \cite{yu2008mixed}}
\label{finger}
\end{figure}

Here, we review the concepts of PSCS  and the details follows from the serial structure. 
First, analog signal  $x(t) $   is time windowed into $M$ pieces, i.e.,:
\begin{equation}
x_m(t)=x(t)w_m(t) \ \ \text{for} \ m=1,...,M; \label{equ13}
\end{equation}  
in which  $x_m(t)$  has a duration  of $T_c$ and $w_m(t)$ is the window function as shown in Fig.~\ref{window}.
 Parallel segmented compressive  sensing consists of a number of parallel 
\emph{fingers}  and each window of signal  is applied to these  fingers  to produce output 
samples. PSCS structure is shown in Fig.~\ref{finger}, wherein  $y_{mN+n} $     represents measuring the $m^{th}$ segment on the $n^{th}$ branch. It should be noted that integrator blocks have the same functionality as the low-pass filter in the serial structure.  
The problem caused by this method is that each segment has incomplete information 
about the signal, and thus all  elements of matrix  $Y_{M\times N}$     should be processed jointly. In other words, we have:
\begin{equation}
Y=[Y_0^T,Y_1^T,..., Y_{M-1}^T], \label{equ14} 
\end{equation}
in which 
\begin{equation}
Y_m=[y_{mN},y_{mN+1},..., y_{mN+N-1}]^T \label{equ15}
\end{equation}
As the last step in CS, we should have a reconstruction matrix. Similar to matrix $V$  in the
serial structure, here we also have a matrix which  elements are the result of inner 
product of sensing random basis  $p(t)$      and expansion basis  $\Psi(t) $      with slightly 
adjustment for segmentation.   

\section{Transmission rate reduction}\label{rate}
So far, we studied the discrete and continuous-time formulations of compressive 
sensing to reduce the sampling rate in ADC devices. Besides the ADC application, compressive sensing can serve other goals such as reducing the transmission rate in battery-powered wireless sensor networks. Wireless sensor networks boast numerous applications ranging from home appliance
control to environmental monitoring and smart healthcare. For instance, they have demonstrated
benefits for intelligent transport systems: monitoring aircraft systems and parameters \cite{chosen},
monitoring wheel bearings on trains \cite{gruden2009reliability}, and connecting sensors, switches, and actuators
inside cars \cite{hashemi2014intra,hashemi2013intra}.  In these applications, energy consumption by sensor nodes is a critical issue as they are typically powered by battery. In most of wireless sensor platforms, radio components of nodes consume most of energy.  Previous experiments and
simulations in \cite{prayati2010modeling} show that energy consumption
by the radio is dominant in wireless sensor
motes. For instance, the radio of
TelosB motes (CC2420) draws 11 mA current to transmit at a power of -10 dBm and 17.4 mA to
transmit at a power of 0 dBm.  However, it should be noted that other
components, such as the processor, consume
energy, and that their energy consumption
depends on the instructions run by the mote.   

Motivated by the power consumption issue, we investigate an application of compressive sensing theory in wireless sensor networks to reduce the rate of transmissions and thus the amount of power consumption. As discussed earlier, compressive sensing  provides a structure to acquire data at a rate proportional to the information rate rather than the rate required by the Nyquist rate. Transmitting data (similar to sampling in the context of ADC devices) with a lower rate can be highly beneficial for battery-powered sensor nodes such as biomedical sensors. 

\begin{figure}[tb]
\centering
\includegraphics[scale=.5]{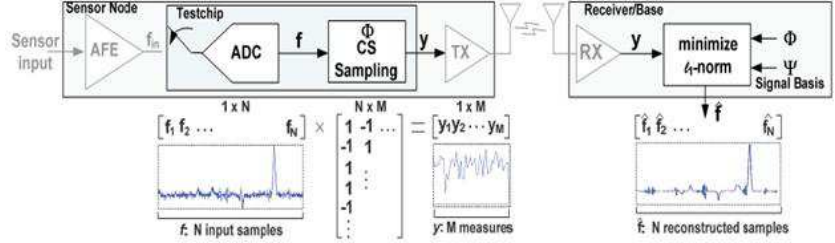}
\caption{Transmission rate reduction with compressive sensing \cite{chen2010signa}}
\label{sensor}
\end{figure}

Within this context, Fig.~\ref{sensor} shows a schematic of sensors with compressive sensing. In this figure, the transmission rate decreases proportional to the rate of compression, and thus it can significantly reduce the actual transmission rate (i.e., without compression) and the power consumption.  
It is worthy to note that the  CS  mechanism  of compression and reconstruction  follows 
the same procedure discussed before, and that in most of wireless sensor networks, the sampling frequency is rarely a limiting factor, which implies that CS can be applied to the digitized outputs of ADC  rather than prior to digitization.

\section{$p$-norm minimization}
\label{idea}
In this section, we propose an enhancement to the compressive sensing framework, with the goal of faster and simpler convergence in the CS reconstruction stage. In the previous section, we discussed about the compressive sensing and that to reconstruct the signal, one should be able to solve the following minimization problem:
\begin{equation}
\begin{split}
\alpha^*=\argmin | \alpha |_1, \\
\text{subject to} \ \ y=\Phi \Psi\alpha.
\end{split} 
\end{equation}
It should be noted that zero-norm can also be used for the purpose of forcing the sparsity. In both cases of zero-norm and one-norm, efficiently solving this minimization problem is critical, for example, for hardware components with limited computation power. However, in this formulation (zero or) one-norm function is not differentiable, and thus gradient-based methods can not be applied.  Using the general definition of $p$-norm, 
$$
|x|_p= \left(\sum x_i ^ p\right) ^ {\frac{1}{p}}
$$
one can see that we can benefit from differentiability by taking the value
of $p$ to be something close to zero and not exactly zero. In this way
based on the above definition we will have a differentiable version of
the zero-norm and so we can use gradient descent methods for solving
the corresponding least squares problem. However, in this case the problem of convexity arises that for $p<1$ the norm is not convex and therefore the problem cannot be solved in a
unique fashion and there will be many convergence issues. Within this context, we propose to use a differentiable and ``smooth" version of the one-norm
and in this way we can benefit both the convexity and the
differentiability. The authors in \cite{lustig2007sparse} use a similar approach for MRI imaging applications, while here we propose to use this algorithm in the context of rate reduction of high-speed ADC devices. 

 \section{Conclusion}\label{Conclusion}
 In this paper, we reviewed the concepts of compressive sensing and its application in
reducing the sampling rate of high-speed ADC devices. Continuous-time formulation 
of CS is studied within the serial and parallel realizations, which, at their core, consist of
three main components: (i) pseudo random sequences, (ii) low pass filter, and 
(iii) low-rate ADC. The idea of reducing  the sampling frequency has various applications such as sampling rate reduction for 60 GHz UWB  communication  and power saving in wireless sensor networks by decreasing the transmission rate.  In this paper, we proposed a tweak to the norm optimization  in the compressive sensing structure in that one-norm minimization is replaced with a $p$-norm minimization for $1<p$ and close enough to $1$. In this case, we can benefit from both convexity and differentiability of the function, and thus the gradient-based methods can be applied. 


\bibliographystyle{ieeetr} 
\bibliography{Summaries}
\end{document}